\documentclass[5p,authoryear]{elsarticle}
\usepackage{graphicx}
\usepackage{txfonts,color,natbib}
\journal{New Astronomy}

\begin{document}
\newcommand{\mps}{m\,s$^{-1}$}
\newcommand{\arcsec}{$^{\prime\prime}$}
\newcommand\apj{{ApJ{ }}}%
\newcommand\apjl{{ApJ{ }}}%
\newcommand\ao{{Appl.~Opt.{ }}}%
\newcommand\aap{{A\&A{ }}}%
\newcommand\mnras{{MNRAS{ }}}%
\newcommand\solphys{{Sol.~Phys.{ }}}%
\newcommand\nat{{Nature{ }}}%

\begin{frontmatter}

  \title{Space--time segmentation method for study of the vertical structure and evolution of solar supergranulation from data provided by local helioseismology}
  
  \author[au]{R.~\v{Z}leb\v{c}\'ik}
  \ead{radek.zlebcik@seznam.cz}
  \author[mps,asu,au]{M.~\v{S}vanda\corref{cor}}
\cortext[cor]{Corresponding author. Address: Max-Planck-Institut f\"ur Sonnensystemforschung, Max-Planck-Strasse 2, 
  D-37191 Katlenburg-Lindau, Germany. Tel.: +49 5556 979 442; fax: +49 5556 979 240.}
  \ead{svanda@mps.mpg.de}
  \author[asu]{M.~Klva\v{n}a}
  \ead{mklvana@asu.cas.cz}
  
  \address[au]{Charles University in Prague, Faculty of Mathematics and Physics, Astronomical Institute, V~Hole\v{s}ovi\v{c}k\'{a}ch 2, CZ-18000 Prague~8, Czech Republic}
  \address[mps]{Max-Planck-Institut f\"ur Sonnensystemforschung, Max-Planck-Strasse 2, 
  D-37191 Katlenburg-Lindau, Germany}
  \address[asu]{Astronomical Institute, Academy of Sciences of the Czech Republic (v.~v.~i.), Fri\v{c}ova 298, CZ-25165 Ond\v{r}ejov, Czech Republic}

  \begin{abstract}
  Solar supergranulation remains a mystery in spite of decades of intensive studies. Most of the papers about supergranulation deal with its surface properties. Local helioseismology provides an opportunity to look below the surface and see the vertical structure of this convective structure. We present a concept of a (3+1)-D segmentation algorithm capable of recognising individual supergranules in a sequence of helioseismic 3-D flow maps. As an example, we applied this method to the state-of-the-art data and derived descriptive statistical properties of segmented supergranules -- typical size of 20--30~Mm, characteristic lifetime of 18.7~hours, and estimated depth of 15--20~Mm. We present preliminary results obtained on the topic of the three-dimensional structure and evolution of supergranulation. The method has a great potential in analysing the better data expected from the helioseismic inversions, which are being developed.
  \end{abstract}
  
  
  \end{frontmatter}
   
\section{Introduction}

Motions of plasma in the solar photosphere have a few characteristic spatial scales \citep[see review by][]{2009LRSP....6....2N}. The convective nature of the solar subsurface layers can be easily seen in solar granulation, convective cells with characteristic scale of 1~Mm and lifetime of $\sim$10~minutes. Exploding granules form a larger apparent structure, the mesogranulation. Weak background magnetic field is not distributed randomly in the photosphere, but concentrated in lane-like structure on the boundaries of supergranules. Larger scales in motions extend further towards larger structures (giant cells, Rossby waves) and end at a differential rotation. Understanding these motions is an important step towards understanding the solar dynamo, a mechanism for origin and redistribution of solar magnetic fields. 

It seems that the solar supergranulation is especially important in forming surface magnetic features. It has been known since 1950s from surface Doppler measurements \citep[e.g.][]{1954MNRAS.114...17H,1962ApJ...135..474L,1964ApJ...140.1120S}. Since then, a vast amount of papers has been published with the topic of supergranulation. It is generally agreed that supergranules are cell-like structures with typical spatial scale of 30~Mm, lifetime of $\sim$24~hours, having predominantly horizontal velocity field with RMS of $\sim$300~\mps{}, and with basically no temperature contrast between the middle of the cell and its boundary. The origin of supergranulation is currently unknown. The classical mechanism invoked to explain the origin of supergranules is the latent heat of the recombination of He$^{2+}$ into He$^+$ at roughly 10~Mm below the photosphere \citep{1985ApJ...288..795G}. In a fluid at rest, such a heat release may trigger  an instability, which can turn into motions at scales comparable to the supergranular scale. Recently, convection structures on supergranular scales have been seen in hydrodynamic simulations of subsurface convection \citep[e.g.][]{2002NCimC..25..523R,2006ESASP.624E..79S,2008AIPC.1043..234U}. 

The development of local helioseismology has made it possible to investigate subsurface layers of the solar convection zone. Time--distance helioseismology relies on analysis of propagation of acoustic waves through the solar interior. The time, in which the wave packets travel from one point to another one, is affected by inhomogeneities along the propagation path. Using various procedures, it is possible to invert the measured travel-time deviations in order to obtain information about the inhomogeneities in the solar interior, e.g. plasma flows. 

A traditional approach is an extension of the principles of geometrical optics, in which rays approximate wave-packets propagation. This approach has been used in many studies and it has been shown that it provides credible results for large-scale perturbations with a reasonable computational efficiency. New inversion methods using more realistic approximations (e.g. first order Born approximation) are being developed. Although the tests using simulated data \citep[e.g.][]{2003ESASP.517..417Z,2007ApJ...659..848Z} demonstrate the credibility of ray-based inversions for the horizontal flow components in the first 5~Mm (and perhaps 10~Mm), independent studies with a different inversion technique \citep[e.g.][]{2008SoPh..251..381J} suggest that already at 4~Mm depth the measured flow field is dominated by noise. This issue is currently being intensively worked on \citep[e.g.][]{2010AA....prepS}.

Flow maps computed by any method of local helioseismology record the structure and amplitude of flows in the certain layer of solar convection zone, given by the averaging kernel also computed by the inversion. The averaging kernel provides important information about the smoothing of the resulting flows. Another important parameter is an accuracy of the results, which is not precisely known in RLS ray-kernel based inversions (it is, however, estimated as a relative error of 10\% in near-surface layers -- \citealt{2001ApJ...561L.229B}). The accuracy is known in e.g. OLA-based inversions \citep[e.g.][]{2008SoPh..251..381J,2010AA....prepS}, which take into account the covariance matrix of travel times and compute the propagation of errors in travel times into the inverted quantity. Helioseismic flow inversions allow for measurements of the flow maps at different depths and therefore provide a 3-D tomographic image of the flows in the solar interior. Therefore, they are suitable for studying the evolution of 3-D structure of convection structures, such as supergranulation.

Three-dimensional inversions of travel times in quiet Sun regions have been shown in past studies, \citep[e.g.][]{1997SoPh..170...63D,1997ASSL..225..241K,2003ESASP.517..417Z,2009SPD....40.0931D}. The usual way of studying the vertical structure of supergranulation is the correlation between the flows at the given depth with the flows on the surface. \cite{1998ESASP.418..581D} showed that in the upper 5~Mm, the flows are correlated positively, then the correlation drops to zero and is negative in the range of 5--8~Mm, suggesting the existence of the ``return flow''. The correlation disappears below 8~Mm, which leads to the depth of the supergranulation of 8~Mm. Similar analysis by \cite{2003ESASP.517..417Z} ended with the estimate of the supergranulation depth of about 15~Mm. \cite{2009NewA...14..429S} compared the large-scale flow from helioseismic data-cubes with the surface flow derived from the tracking of supergranules in MDI Dopplergrams with a conclusion that the large-scale velocity field does not change much in upper 10~Mm depth and supergranules can be therefore treated as objects carried by the underlying flow.

The main goal of our study is the development of an efficient code to study evolution of the three-dimensional structure of supergranulation. The paper is organised in a following way: in Section~2 we explain the segmentation method used to isolate individual 3-D supergranule and their histories in the sequence of helioseismic data-cubes. In Section 3 we briefly describe the helioseismic data we have at our disposal and the results we obtained after application of the described method. We conclude in Section 4 and discuss a future perspectives of the presented study.

\section{Segmentation method}

In order to isolate the supergranules and their histories in the sequence of 3-D flow maps we must first define a supergranular cell. It has been shown in earlier works \citep[e.g.][]{2001ESASP.464..577B,2003Natur.421...43G,2008SoPh..251..417H} that a convenient way to identify the individual convective cells, is the use of the horizontal divergence $\nabla_h \cdot \mathbf{v}$, 
\begin{equation}
\nabla_h \cdot \mathbf{v} = \frac{\partial v_x}{\partial x} + \frac{\partial v_y}{\partial y}\ ,
\end{equation}
where $v_x$ and $v_y$ are horizontal components of the flow field. On the surface, $\nabla_h \cdot \mathbf{v}$ mimics the vertical component $v_z$: the regions of a positive horizontal divergence are usually regions of up-flows, while regions with a negative horizontal divergence depict down-flows. It would be wiser to use $v_z$ directly in identifying the supergranules, however it is not possible in current inversions, because a weak signal of $v_z$ is buried in noise and systematic errors, such as a crosstalk. 

A supergranule in a flow map near the solar surface is defined as a compact region of the positive divergence surrounded by the lanes of the negative divergence. In order to avoid the noise, thresholding and limitations on the detected structure size are necessary. The segmentation algorithm is, however, independent of the supergranule definition and in principle should be usable for any reasonable definition. 

\begin{figure}[!ht]
\includegraphics[width=0.5\textwidth]{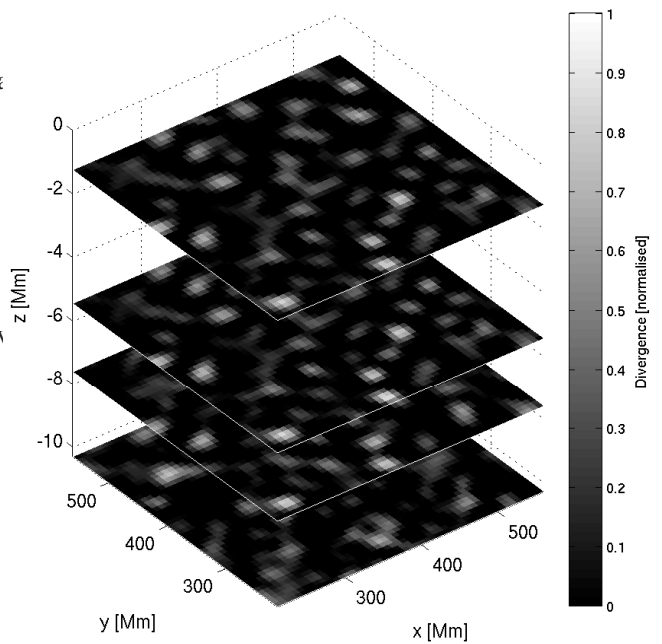}
\caption{An example of the horizontal divergence (positive part only, normalised to the maximum value) of the supergranular flow measured by time--distance helioseismology at four different depths. The data were smoothed by the low-pass Butterworth filter in order to suppress noise. We see that in the first three depths the pattern remains largely unchanged, i.e. depicting compact supergranular cells, while in the last displayed layer the pattern is different (slightly negatively correlated to the previous ones). The layer at the depth of 7.6~Mm is a transition layer between these two. The aim of the segmentation routine is to find and connect the corresponding areas with high value of positive horizontal divergence into compact three-dimensional object and track them in the consequent 3-D horizontal divergence data-cubes. }
\label{fig:3d-divergence}
\end{figure}

In our case, we smooth all divergence maps in $k$-space by a fourth-order low-pass Butterworth filter in order to supress the structures that are significantly smaller than the expected size of supergranules. The threshold for the regions, which are considered as candidate supergranules, is selected as a $\epsilon$-multiple of the RMS-divergence at the given time and depth-layer. As a result, the input of the segmentation routine is a set of consequent three-dimensional divergence maps, in which we keep smoothed regions with the horizontal divergence above the threshold, surrounded by the lane-like regions having zero value (cf. Fig.~\ref{fig:3d-divergence}). The threshold parameter $\epsilon$ is subject to the tuning, but remains constant for all times and depths, so that it takes into account possible systematic change in the flow amplitude with time and depth. Here $\epsilon=1.5$.

The segmentation algorithm is based on the feature tracking approach, where we track supergranules from the surface down in depth and forward in time. In ``(3+1)-D segmentantion'' number 3 stands for the spatial dimensions and 1 for time. We start from the near-surface layer at the starting time $t_0$, which corresponds to the beginning of the sequence of data-cubes. In the processed divergence map we label each continuous region with divergence values above the threshold by a unique tag -- these are possible supergranules. These supergranular candidates are then tracked down to the deeper layers at the same time, assuming that the location of the same cell at different depths does not change significantly. In this way, we select three-dimensional compact objects (3-D supergranules at time $t_0$). 

In the next step, we move to the consequent 3-D divergence data-cube at time $t_1=t_0+\Delta t$. We assume that during the time step $\Delta t$, which divides two contiquous data-cubes (here $\Delta t=8$~hours), the supergranules did not move significantly, i.e. that supergranular candidates at a similar position represent the same supergranule at two consequent times. Given the diferential rotation rate in the solar photosphere, this assumption is very reasonable, the estimated shift is 2--4 pixels with resolution of 1.4~Mm. In the near-surface layer, we search for supergranules known from the previous time-step. If some of them are not identified, it is assumed that they are diminished and their history ends here. Newly detected supergranules, without an appearance in the previous time-step, are labeled with the unique tag. The search is then extended into the depth domain. As a result, we obtain the first set of three-dimensional supergranules with parts of their histories, and a new set of three-dimensional supergranules, which do not have a history yet, because they have been seen for the first time. We then move to the consequent data-cubes at times $t_2=t_1+\Delta t$, $t_3=t_2+\Delta t$,~\dots, and repeat the algorithm again, until we reach the end of the continuous sequence of data-cubes. The procedure results in a set of three-dimensional supergranules and their histories, describing their location in space and time. 

Physically it is possible that the supergranule splits in two, or that two supergranules merge into one. If this happens, the split or merged supergranules are labeled as completely new ones, ending the life of their predecesors. Supergranules not having a complete history, i.e. being on the edge of the data-cube, are removed from the sample. 

\section{Results}

\subsection{Data}

We have developed and tested the method on data-sets available today, which consist of a series of 3-D flow data-cubes from local helioseismology (see example in Fig.~\ref{fig:3d-divergence}). These flow data-cubes were obtained by ray-kernel based time--distance helioseismology \citep{1993Natur.362..430D,1997ASSL..225..241K} using phase-speed filters and RLS fitting, and are considered state-of-the-art. The data we have at our disposal were used in previous studies \citep[e.g.][]{2006SSRv..124....1K,2009NewA...14..429S} and consist of 46 flow-map sets in the vicinity of the active region NOAA~9393/9433. This sequence of three-dimensional flow maps was recorded in March and April 2001. To obtain reasonable signal-to-noise ratio, each of these maps is averaged over 8~hours and also in spatial coordinates, so that the effective resolution of the results is 5.6~Mm. The data-cube contain two temporal discontinuities, when the active region was situated on the far side of the Sun. 

The data were originally computed for studies of the depth structure and evolution of a large active region. The presence of the magnetic field affects convection and also the propagation of seismic waves, therefore magnetic regions are not suitable for studies of supergranulation. For our purpose, we use only quiet Sun regions around the magnetic areas. This selection limits the effective field-of-view and also the number of observed supergranules that can be used for statistical studies. As we intend to develop the method of (3+1)-D segmentation on available data and use it for serious research using better data available in the future, we do not see the relatively small amount of supergranules as a serious drawback. 

\subsubsection{Data testing}

\begin{figure}
\includegraphics[width=0.5\textwidth]{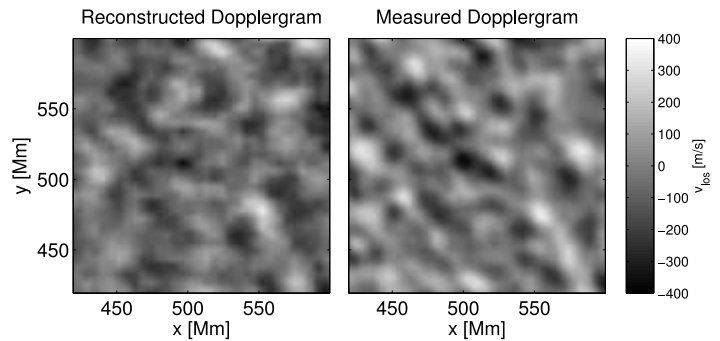}
\caption{A comparison of the real measured Dopplergram and the synthetic one, constructed from the time--distance data. We see that the match is not perfect, however, still reasonable. Correlation coefficient in this example is 0.37.}
\label{fig:dopplergrams}
\end{figure}

In order to test the credibility of the time--distance data we have at our disposal, we constructed the synthetic surface Dopplergrams and compared them with the observed ones. One has to keep in mind that we do not compare exactly the same features: synthetic Dopplergrams are constructed by taking the line-of-sight projection of the horizontal (i.e. $v_x/v_y$) velocity field representing flows at the depth of 0.77--1.77~Mm. We do not have pure surface helioseismic measurements at our disposal, because the surface $f$-mode is filtered out during the application of time--distance helioseismology used for the inversion. Moreover, vertical $v_z$ component is unreliable to use \citep{2007ApJ...659..848Z}. Therefore, we cannot expect a perfect match between the reconstructed and measured Dopplergrams. 

Nevertheless, as we see in Fig.~\ref{fig:dopplergrams}, we find many similarities. The correlation coefficient is typically 0.4 and it is higher at around 55 degrees from the disc centre. Based on the direct comparison with the surface measurements, we may estimate that the accuracy of helioseismic data lies between 10 and 20\% of relative error.

\subsection{(3+1)-D segmentation}

As already mentioned, we define the supergranule as a compact region of the positive horizontal divergence in the near-surface layers. Therefore, we first use horizontal components of the measured plasma flow for computation of the horizontal divergence. Resulting (3+1)-dimensional divergence data-cubes are smoothed by a fourth-order low-pass Butterworth filter with a cut-off of 0.034~Mm$^{-1}$ in order to suppress the random noise in the data and enhance the structures on supergranular scales. 

Following that, the segmentation algorithm described in Section~2 is applied. We recognised 1212 supergranular cells and their histories. In the sample, short living cells having volume less then 300~Mm$^3$ are designated as noise and removed from the set. This value is roughly one tenth of the expected volume of supergranular cell having 30~Mm in diameter and depth of 5~Mm. 

We investigated the resulting set of supergranules to confirm that its statistical properties are consistent with literature. 

\begin{figure}
\includegraphics[width=0.4\textwidth]{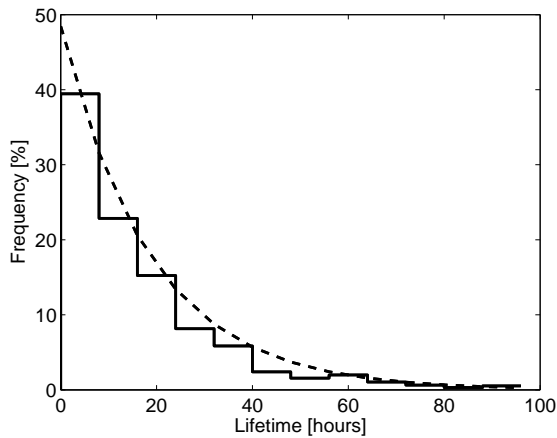}
\caption{A histogram of lifetimes of supergranules detected by the segmentation method. A fit following an exponential decay is over-plotted with an expectation lifetime of 18.7~hours.}
\label{fig:histogram}
\end{figure}

A histogram of the lifetimes of supergranules in the sample is displayed in Fig.~\ref{fig:histogram}. An exponential fit of the form
\begin{equation}
f(T) \sim \exp{\left[ -\frac{T}{\tau} \right]}
\end{equation}
provides an expectation lifetime of $\tau=18.7$~hours, which is in agreement with e.g. \cite{2004ApJ...616.1242D} or \cite{2004SoPh..221...23D}. From our analysis, it becomes clear that supergranules with higher values of the horizontal divergence in near-surface layers depict longer lifetime, in agreement with \cite{2008SoPh..251..417H}. The dependence is almost linear for supergranules having maximum horizontal divergence values 10--60~$\mu s^{-1}$, for supergranules having maximum horizontal divergence larger than 60~$\mu s^{-1}$ the lifetime remains almost constant ($\sim$30~hours), again with an agreement with \cite{2008SoPh..251..417H}. The typical size of the detected supergranular cells is 20--30~Mm. The lifetime of the cells with this size is around 25~hours. 

Having supergranules defined as three-dimensional objects in the time sequence makes it possible to study their evolution. We do not observe significant rise or decline of the typical supergranule during its lifetime. The gravity centres of the horizontal divergence regions remain at the depth of 3.2~Mm over almost the whole life-span. We measure an insignificant average rise of the gravity centre in the growing phase of the supergranules by roughly 0.2~Mm with the speed of 3~\mps{}, both values are beyond the precision of the used data. In some supergranular histories, we detected a initial rise of the region with the maximum divergence with the speed of (300$\pm$100)~\mps{}. These observations unfortunately strongly depend on the quality of the used data, which might be in doubt in our case. Therefore in the present analysis, we are far from making conclusion about the physics of supergranules. 

\begin{figure}
\includegraphics[width=0.45\textwidth]{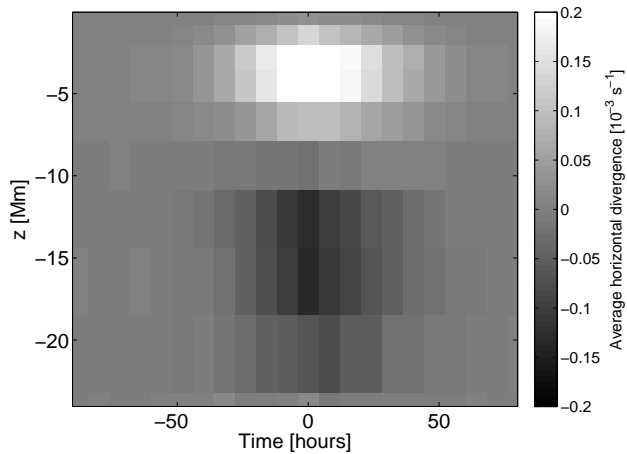}
\caption{The average supergranule evolution in time and depth, averaged over 215 supergranules with the lifetime longer than 24~hours. Shades of grey depict the average value of the horizontal divergence in the supergranular cell in the given time and depth. }
\label{fig:averageSG}
\end{figure}

The evolution of the average supergranule in time--depth space is displayed in Fig.~\ref{fig:averageSG}. This is a result of averaging over 215 supergranules which survived for more than 24 hours. All supergranules were co-aligned in time to the moment, when the maximum of the divergence in the near-surface layer of 0.77--1.77~Mm were established. The characteristic horizontal divergence used in this plot was calculated as a horizontal average of the horizontal divergence in the supergranular cell at the given time and depth. 

The overall picture on the plot again depends highly on the accuracy of the helioseismic data, especially whether the data are free of systematic errors. Nevertheless, current results confirm that supergranules are of convective origin. They consist of the region with positive divergence in the sub-surface layers and the region of the negative divergence, the region of the ``return flow'', extending deeper down. The upper ``outflow'' region evolves slowly in the intensity over the course of supergranular life, but remains confined in the depth till $\sim$6~Mm. The ``inflow'' region seems to extend almost twice as deep at the maximum phase than during the growing and decaying phases. Based on the displayed results we estimate the depth of supergranulation of $\sim$15~Mm at the growing and decaying phases, and $\sim$20~Mm at the maximum phase. 

With a certain amount of phantasy we could say that the average supergranule starts to form 50~hours before its maximum at the depth of $\sim$7~Mm and extends towards the surface and deeper down as the time progresses. The evolution in time is slightly asymmetrical showing insignificantly slower decay compared to the formation. In the average picture, we do not see signs of structural rise or decline in depth. 

\section{Conclusions}

We have developed and tested the method of (3+1)-D segmentation of supergranules as evolving three-dimensional objects in the series of helioseismic three-dimensional flow maps. The segmented supergranules display surface statistical properties that are in agreement with widely accepted parameters, e.g., typical size of 20--30~Mm, characteristic lifetime of 18.7~hours, and estimated depth of 15--20~Mm. We made an attempt to investigate the evolution of the average supergranule in time and depth, which could confirm the convective nature of supergranules. The results do not address the issue of tendency of supergranules to support waves \citep{2003Natur.421...43G}. The wave-like properties manifest as apparent motions of supergranular structure, which are not decomposed when tracking individual supergranular cells in space during the process of segmentation.

From our analysis it seems that supergranules extend deeper at the maximum phase of their evolution than during the phases of growth and decay and have slightly asymmetric evolution in time with decay slower than growth. 

These results are, unfortunately, sensitive to the quality of the data used for segmentation. We are well aware of papers which doubt ray-kernel based inversions showing that wave-effects have to be taken into account \citep[e.g.][]{2000SoPh..192..231J,2000SoPh..192..193B} as the wavelength of seismic waves is comparable with the size of inhomogeneities in the Sun. It is therefore likely that wave-based kernels could yield depth inversion results that are qualitatively different from the ones based on ray kernels. However, the data used in this study are still considered state-of-the-art and provide us with sequences of data-cubes suitable for developing and testing of the described method. Born-kernel based time--distance inversions for flows are being developed \citep[e.g.][]{2008SoPh..251..381J,2010AA....prepS}, especially in the frame of observations coming from Solar Dynamics Observatory. It will still take some time before the higher-quality flow maps suitable for (3+1)-D segmentation will be available. We are looking forward to this data and believe that our method applied to these data-sets will provide a deeper insight into the 3-D structure and evolution of supergranulation. 

\section*{Acknowledgement}
We thank Dr. Alexander G. Kosovichev, Stanford University, USA, for providing us with the data. This work was supported by the Grant Agency of Academy of Sciences of the Czech Republic under grant IAA30030808, M.~\v{S} additionally by ESA-PECS under grant 98030. The Astronomical Institute of ASCR is working on the Research project AV0Z10030501 (Academy of Sciences of CR), the Astronomical Institute of Charles University on the Research program MSM0021620860 (Ministry of Education of CR). SOHO is a project of international cooperation between ESA and NASA.




\end{document}